# Photonic Framework: Cues To Decode Quantum Mechanics

O. Tapia
Chemistry Ångström, Uppsala University, Sweden

The photonic framework [1,2] offers clues leading to a possible update of quantum mechanics' foundations yet preserving its mathematical grounds. Reconsideration of quantum theoretic measurement theory warrants removal of ideological prejudices brought in by von Neumann´s Copenhagen interpretation.[3] Conditions required to bridge abstract to laboratory space formulations of QM permit underscoring fundamental differences to the standard view; entanglement and sustainment concepts help in this task. Entanglement events integrate probing devices leading to q-measurements seen as physical processes. Representation elements based on classical physics pictures are removed. A non-representational framework stands up as a result.

**Introduction**
The possibility for a re-foundation of QM opens paths to a viable program provided quantum states (q-states) do not represent objects (entities, molecules); [1,2] viz., to leave behind for instance handling of nuclear positions (Born-Oppenheimer models) as parameters and instead introduce a spectral/spectroscopic mode.[4,5] Recasting helps formulations in abstract space of a theory able now to incorporate quantum entanglement features. Thereafter it lets open grounds for examining links between abstract to laboratory space via suitable semi-classic models that open access to possibility spaces rather than probability ones. In this context Löwdin formalism [5] can with no trouble be recovered once different degrees of freedom are plainly admitted via quantum numbers and one uses them as base states labels in abstract Hilbert space. Thereafter links obtain by explicit incorporation of photon-states rooted in Fock spaces.

In the photonic approach [1,2] the elementary material constituents *sustain* quantum states; thus, electrons and nuclei altogether acting as elementary materiality actively *support* electronuclear (EN) q-states; such q-states would not represent this materiality either as particles or as waves (in a classical physics sense) and/or molecules, yet q-states as such mediate interactions, with for example external sources leading to characteristic responses (spectra).

Particle occupancy concept is thus taken away because theory has not representational aims: electrons do not occupy particular basis states; this idea may only be useful in semi-classic frameworks. This perceptive change opens handling of mathematical symbols on quite general grounds independent from ontic/ontological remnants or idealist/materialist tenets. q-States, in infinite numbers, can hence be ordered under different spectral characteristics and, most importantly, *the same* elementary materials would sustain them all. In this context the linear superposition principle finds a natural place (not a postulate).

One net gain in going over a non-representational approach is that most difficulties encountered in for example developing quantum electrodynamics (QED)[6] are bypassed. Wedding matter q-states to Fock *photon-number* base states can be done within fully abstract schemes. Sustainment of photon base states (laboratory space) rests upon radiation energy. *It is energy exchange that is quantized*. Yet, the quantum is neither a particle nor a wave in any classical physics sense.

This paper further elaborates on some quantum physical characteristics starting from a semi-classic scheme to fix ideas. Abstract schemes are naturally introduced as shown by von Neumann´s mathematical work,[3] though definitely subtracting interpretations such as collapses of the wave function etc. Thereafter, abstract and laboratory spaces are constructed helped by introduction of entanglement concept and links between them emphasized. The way this new approach provides a substitutive foundation leads to a sort of decoding to QM. Suitable (adapted) semi-classic models employ a non-representational approach to q-states while interactions with external sources do incorporate geometry elements found in laboratory space.

**Beyond standard semi-classic models**
Semi-classic models, as developed so far, belong to laboratory quantum physics extensible to chemistry and biology; they have helped elaborate computing replicas in QED and molecular Quantum Mechanics underlying applications to Quantum Technology designs.[7-12] Their *usefulness* is certain although their quantum grounds precarious. Also, without inclusion of entanglement, this model class sidesteps q-events that now would become keys to help introduce bridges (gangplanks) from abstract to laboratory domains. Bridges, yes, but not representations; focus is now on interactions that open the system to new possibilities. Basis states for entanglement contribute supplementary elements to stitch independent partite system models as shown below.

q-Events find expression in q-interactions, particularly apparent for q-systems under probe and laboratory measuring (probing) devices; q-events communicate as it were via (in principle detectable) energy and angular momenta q-*transfers*, thereby opening windows (fences) between abstract and laboratory worlds. Recording q-events is thus a fixing step in measurement procedures, e.g.: traces in Wilson cloud camera and/or images rooted therein, etc.

To achieve a particular model, material purview must both be taken as quantized (q-fields). In this manner one gets Fock basis elements $\{|n_{\omega j}>\}_{\omega j}$ *related to* radiation q-states and on the other hand matter sustained basis sets $\{|ik(i)>\}_{i,k}$.[4,5] Laboratory (physical) Fock basis are elements *sustained* by radiation energy of appropriate frequency, i.e., an executive presence of electromagnetic (EM) energy *warranting functionality*. Once elementary constituents are chosen, a large array of chemical species may relate to particular sets of amplitudes (q-states) that typify quantum possibility sources. One thus enters a different world (one of possibilities), akin to a mathematic world. So called excited states are automatically counted in.

Now, let us shift theory's grounds beyond standard semi-classic levels.[9,12-15] As R.G. Newton asserts[13] a state vector does not describe the system itself and then, it is the idea of representation that falls off; see also refs.[9,14] Amplitudes not only serve to calculate probabilities (if at all) but most importantly intervene in q-interaction presentations between matter sustained q-state and those associated to probing space. [14] Probabilities would play a subsidiary role not required to explain a measurement process yet offers a handy language for evocative purposes more related to a classical world.

A theoretic measurement calls for entanglement/disentanglement that, as a resource, bridges probe/probing edges thereby compelling introduction of entanglement events. One takes the probe/probing edges first as independent partites and measurement imposes construction of an interacting system.

Importantly, simultaneous variation in both q-fields implies non-separability that here it is implemented with extra base states prompting for such q-interactions, a minimal subspace basis elements is thence subsumed into the matter sustained basis set.

Energy/angular momentum reshuffling sensed at recording stage, where partite states can be related to separated probe/probing elements; and as result of the measuring procedure their states are concomitantly changed.

Now, it follows from the above that classical physics is not central to a measurement event and this to the extent that, even in standard QM, local realism can safely be set aside also. [10-16] Linking abstract to laboratory spaces is henceforth the target.

The present paper partially takes ground from the Quantum Electrodynamics ideas advanced by Dirac, Fock and Podolsky, [17] although taken as a resource. A non-representational mode is implied.

**Constructing q-states**
Consider a Hamiltonian operator $H \rightarrow \sum_i \varepsilon_i |i\rangle\langle i|$; with $\{\varepsilon_i\}$ theirs eigenvalues and $\{|i\rangle\}$ taken as base elements of a Hilbert space, the q-states take on the forms:

$$|\text{q-state}\rangle \rightarrow (C_1\ C_2\ C_3\ldots C_k\ldots) \rightarrow \sum_i C_i |i\rangle \qquad (1)$$

The connection relation ($\rightarrow$) emphasized; the commutator $[H, |i\rangle\otimes\langle j|]$ leads to matrix operator elements: $H|i\rangle\otimes\langle j| - |i\rangle\otimes\langle j|H$. Applying this commutator to a $|\text{q-state}\rangle$ one picks up an information $(\varepsilon_i-\varepsilon_j)\ |i\rangle\otimes \langle j|\text{q-state}\rangle$ embodying a complex number and, in a way of speaking, it extracts up amplitude at j-th base state $C_j \rightarrow \langle j|\text{q-state}\rangle$ thereby showing its modulating role in response to a possible transition valued as $(\varepsilon_i-\varepsilon_j)$. Base state $|\varepsilon_i\rangle$ plays a role of target state and $|\varepsilon_j\rangle$ a root state. The complex number effect now the target state $|i\rangle$. It is in this sense that amplitudes may play a direct physical role: namely, as modulators.

The case to be scrutinized corresponds to that of a photon field *mediating* exchange of one radiation energy quantum. Fock space completes the scope of the model (see below). The energy quantum relates two energy level terms with Bohr's map: $(\varepsilon_i-\varepsilon_j) \rightarrow h\nu_{ij}$ the frequency field indexed by two integers makes part of a matrix form. This later $h\nu_{ij}$ is translated into: "quantity of EM-energy measured thence by the value at one radiation frequency". Thus, $h\nu_{ij} \langle i|i\rangle\otimes \langle j|\text{q-state}\rangle$ stands for a quantum transfer between the material and radiation fields. At this point one can sense information that may be lost due to the replacement of the matrix form $\nu_{ij}$ by one frequency value, say ν corresponding to radiation energy; the following corresponds to a especial semi-classic scheme thus it is not necessarily a "law of nature" but rather a q-event.

What the q-event stands for is an apparent energy swap between a photon q-field and a matter-sustained q-field. Note that the amplitude may vary in a continuous manner.

Bohr's map relates to energy level differences and in this sense one would map to a measurable quantity such as an energy quantum. *Eigen-values are not "observables" in themselves*. [1-2,9,14] And the symbol hν, *by convention*, stands as value of an energy quantum, h being Planck´s constant. This later statement is another important departure from the one found in standard QM where jumps play a central role.

Eigen value is thus *a resource* offered by quantum formalisms. Yet, for a given q-state at probe it is apparent that exciting from a given base state (e.g., a root state) *requires the particular amplitude of the q-state to be different from zero, eventually a function of time*. The link to the elementary materiality is unspoken (understood).

**Basic Photonic Scheme: energy/momentum swaps via entanglements**
In the simplest model mentioned above, two classes of base states enter the report: 1) matter-sustained EN base states $\{|ik(i)>\}$ with label indicating principal q-number i and k(i) subsidiary q-numbers; 2) Fock photon-number base states elements $|n_{\omega j}>$; $\omega_j$ signals a given radiation frequency taken as *label*; there is no physical model assigned to this frequency, nothing "vibrating" as it were. This model is adequate for non-interacting partites to be seen initially as definite partites logical sums states: $\{|ik(i)>\} \oplus \{|n_{\omega j}>\}$.

To imply possible q-interaction in abstract space the direct product form would in part do: viz., $|ik(i)>\otimes|n_\omega>$. For entanglement interaction take $|ik(i);1_\omega>\otimes|n-1_\omega>$ (as a possibility). The symbol $|ik(i);1_\omega>$ corresponds to *matter-radiation q-entanglement* i-th basis element as the possible root state; this stands as a prototype possibility of quantum effect with no counterpart in classical physics world. Yet a simple direct product would not suffice. The minimal subspace $|ik(i)>\otimes|n_\omega>$ and $|ik(i);1_\omega>\otimes|n-1_\omega>$ shows effective 2-dimension resembling a q-bit. These kinds of two-basis set are now taken as information elements with a q-bit surged into a larger q-state form; for a *one-photon case* relevant non-separable base states would read as: [2,3]

$$(\ldots|ik(i)>\otimes|1_\omega>\ldots|ik(i);1_\omega>\otimes|0_\omega>\ldots|i'k'(i');0_\omega>\otimes|0_\omega>\ldots|i'k'(i')>\otimes|0_\omega>\ldots)^T \to <\text{basis}| \qquad (2)$$

The basis vector would cover relevant possibility sources related to the q-interacting system and, in this sense the initial logical sum of partite states would have to be link with another one displaying entanglement as a natural possibility; it is the link to be discussed later on.

Firstly, note the special entangled base state $|i'k'(i');0_\omega>\otimes|0_\omega>$, where the label i' for instance indicates an energy level matching the one for $|ik(i);1_\omega>\otimes|0_\omega>$ so that the space engaged would be larger that the one commonly used in the particle model approach. Note the effective 4-set:

$$|ik(i)>\otimes|1_\omega>; \; |ik(i);1_\omega>\otimes|0_\omega>; \; |i'k'(i');0_\omega>\otimes|0_\omega>; \; |i'k'(i')>\otimes|0_\omega>. \qquad (2')$$

These four generic levels are by hypothesis energy degenerate. Reading their labels they point to possible energy reshufflings processes that may be coupled to frequencies ranging from radio to microwaves with corresponding low-energy quanta related to k(i) and k´(i`) energy ladders; such proposal does contain basis states signaling possibilities only, so that to complete the analysis there is need for specific q-states.

A particular q-state would come out as row vector assembling labeled complex numbers (amplitudes) *ordered* according to the sequence found in (2)-|basis>; it is using basis labels for the amplitudes that one outlines particular abstract q-states via information content (numeric values for amplitudes and labels), e.g.:

$$(\ldots C_{ik(i) \otimes 1_\omega} \ldots C_{ik(i); 1_\omega \otimes 0_\omega} \ldots C_{i'k'(i') \otimes 0_{\omega'}} \ldots C_{i'k'(i'); 0_\omega \otimes 0_{\omega'}} \ldots) \rightarrow |q\text{-state}> \quad (3)$$

Here, non-zero amplitudes modulate interactions while zero-valued ones *keep the set ordered*.[3,4] This latter factor ignored in standard models for the simple reason that it ignores the "messenger" and obviously the "message".

A particular q-state can be seen as *implicit* scalar product: <basis|q-state>$_\aleph$, where $\aleph$-sign is there to prevent taking a simple (finite) sum only; thus, keeping in mind this caveat, <basis |q-state> stands for full the scalar product, <(2)|(3)>; not a finite sum of terms and the caveat $\aleph$-sign from now on is no longer necessary.

Physical q-processes would be sensed when those initially zero-valued amplitude start displaying non-zero values due to interactions thereby opening to probe. Thus, at a laboratory level such new non-zero amplitudes would permit sensing such changes (using appropriate probes). There will be no need for "collapses" characteristic for the standard QM theory. *What ought to be probed is the whole q-state always*.

Besides, *basis vectors* being an information resource they remain *fix once a particular model is chosen*. They are organized to outline possibilities associated to particular materiality yet conserving not only its abstract character but also the predictive power associated to the possibility fields the basis set stand for.

All combined basis terms once put in (2) conform non-separable basis grounding possibilities associated to the quantum system that cannot be handled separately (independently); changes in q-states would happen via amplitude transformations engaging states like that in eq.(3).

No partitioning included in the matter scheme because it is unnecessary. Partite subsystems with their spectral responses sharing the same total number of elementary material elements are linked (if necessary) via supernumerary entangled basis states; [2,3,9,14] supernumerary states with respect to isolated matter sustained basis elements.

Links to laboratory domain will bring possible chemical species identifiable by spectral responses eventually opened to recording at laboratory level.

The elements like |ik(i)>⊗|1$_\omega$> found in q-state-(2) share a common origin (I-frame) and the same I-frame with all remaining basis elements. Otherwise it would stand for

independent photon source and matter location situation to be considered later on (partite states).

On the other hand, the term $|ik(i);1_\omega\rangle \otimes |0_\omega\rangle$ signposts also photon number depletion and portrays *entanglement of photon and matter field states;* accordingly there is no free quantum of radiation energy available that could be given back if one were to startup probing at this level. This basis term gathers information as a "photon-dressed" state (*photon-matter entangled state*); and it contributes to an *executive* presence to entanglement.

**Reading q-states: possible (virtual) processes**

Most of the situations found correspond to sets of amplitudes values and the way they may change. Consider a q-state amplitude say from $C_{ik(i)\otimes 1_\omega}=1$, $C_{i'k'(i');0_\omega \otimes 0_{\omega'}}=0$ changes into $C_{ik(i)\otimes 1_\omega}=0$, $C_{i'k'(i');0_\omega}=1$, all other implicit amplitudes being unchanged; one gets a state of photon/matter entanglement relating not to objects but q-state mutation possibly prompted by a q-process (not explicitly given now) leading to *amplitudes relocations*. Consider then the change of q-state:

From: $(\ldots 1_{ik(i)\otimes 1_\omega} \ldots 0_{i+1k(i');\otimes 0_\omega} \ldots 0_{i+1k'(i')\otimes 0_\omega} \ldots 0_{i'k'(i')\otimes 0_\omega} \ldots)$ (4a)

To: $(\ldots 0_{ik(i)\otimes 1_\omega} \ldots 1_{i+1k(i');\otimes 0_\omega} \ldots 0_{i+1k'(i')\otimes 0_\omega} \ldots 0_{i'k'(i')\otimes 0_\omega} \ldots)$ (4b)

Taken together (4a) and (4b) define a one-photon ingoing channel seen from matter elements viewpoint, i.e. from laboratory viewpoint: 1-photon state becomes entangled and the excited state i+1 activated with simultaneous depletion of the photon number. This talk results from reading the labels so that it represents information exchange.

Information flows; the photon field would sense an information change once the initial non-interacting partites were bring together (see below for more details).

Materiality sustaining these q-states is necessarily *conserved* and never engaged in "filling" any particular energy eigenvalue. The term $1_{ik(i);1_\omega \otimes 0_\omega}$ tells that amplitude now prompts a possible excited electronic state (i+1)k(i') decorated with a label for entanglement with a *vacuum photon field*. Whatever the experimenter does will be recorded by a corresponding vacuum base state. Thus vacuum states can carry endless information (certainly not energy!).

A second label $0_{i'k'(i')\otimes 0_{\omega'}\otimes 0_\omega}$ carry information that would open possibilities for a second photon interaction so $1_{(i+1)k(i');\otimes 0_\omega}$ should read now $1_{(i+1)k(i');\otimes 0_\omega \otimes 1_{\omega'}}$ as for example activating a domain with larger angular momentum a case to be found for singlet-triplet spin states. This is examined later on in detail.

Now, take the opposite direction, namely, from (4b) to (4a) one gets information that a *possible* emission mode might be open, but the putative emitted state belongs to a limit that does not have its proper place in the space used to introduce $\langle(2)|(3)\rangle$ above. This simply means that abstract theory has to be supplemented to describe q-events, as one would expect. Old quantum mechanics completeness claims are not granted.

Once this caveat is understood the amplitude at entangled base state in (4b) can relate to q-state (4b') or (4c) for instance:

$$(\ldots 0_{i-1k(i-1)\otimes 1_\omega}\ldots 0_{i-1k(i-1);1_\omega}\ldots 0_{ik(i)\otimes 0_\omega}\ldots 1_{ik(i);\otimes 0_\omega}\ldots 0_{i+1k'(i+1);0_\omega\otimes 0_{\omega'}}\ldots) \quad (4b')$$

In eq.(4b') the entrance (emission) channel is closed; i.e., $C_{i-1k(i-1)\otimes 1_\omega}=0$. From here ($1_{ik(i);\otimes 0_\omega}$) activation leading to state (4c) may result via coherent states, though for simplicity sake we show a pure state to introduce a language element:

$$(\ldots 0_{i-1k(i-1)\otimes 1_{\omega'}}\ldots 0_{i-1k(i-1);0_{\omega'}}\ldots 0_{ik(i)\otimes 0_\omega}\ldots 0_{ik(i);\otimes 0_\omega}\ldots 1_{(i+1)k'(i+1);0_\omega\,0_{\omega'}}\ldots) \quad (4c)$$

This latter vector may stand as a triplet excited state ($T_1$) *entangled* to both *ω- and ω'- vacuum base states*. In this manner a "quantum trail" is signified, e.g. label $_{;0_\omega\,0_{\omega'}}$ two light quanta have been expended at different points.

From a semi-classic stand the possibility of a photon state "absorption" (4c) starting from (4a) is apparent. Possibilities for having registered histories are brought in; the use of laser sources help driving quantum changes; physical examples are analyzed in Ref. [14,21]. Before discussing in detail how to "open" a spin triplet state let us examine linking systems corresponding to abstract and laboratory spaces formalisms first.

**Linking systems: abstract - laboratory spaces**
In terms of information data a bridge can be identified connecting laboratory to abstract domains. Special relativity theory (SRT) framework let I-frames in and, via these open a connection (sort of gangplank) between these domains. Importantly, the I-frame sustains configuration spaces that for the present case the dimension defined by *the number* of classical degrees of freedom i.e. a dimensionless number. Two I-frames help define relative origins and orientation required by SRT. Abstract configuration space: $\mathbf{x} \to (\mathbf{x}_1,\ldots,\mathbf{x}_n)$ would collect ordered *information* on 3n degrees of freedom that can eventually be linked to classical physics. The introduction of time and space lies on mathematical grounds. There is no absolute space or time.

A wave function $\Psi(\mathbf{x}_1,\ldots,\mathbf{x}_n)$ comes out as scalar mapping $\langle\mathbf{x}_1,\ldots,\mathbf{x}_n|\Psi\rangle$; this links to a complex function over a real multivalued domain, n-tuple. The I-frame opens such a possibility by facilitating a definition of wave functions; these are elements of a Hilbert space over square integrable (complex) functions. Of course, the set $\{|\Psi\rangle\}$ sustains another abstract Hilbert space over the complex number field as well.

At this point let quest after links from abstract to laboratory space. Both spaces are not commensurable and, resorting to semi-classic frameworks, connections can be constructed helped with configuration space concept and wavefunction $\langle\mathbf{x}_1,\ldots,\mathbf{x}_n|\Psi\rangle$ that one implements as first step leading to connections. From its very inception it cannot be mixed up with an object (entity) and cannot represent a molecule for example.

There is need of a *measure space* to proceed with the integrations that must be implemented; and not only this, geometry concepts would also accompany a number of mathematic operations. Yet again the wave function is never to be taken as an object of laboratory space, it is not "real".

Invariance groups associated to I-frames translations and rotations enter the stage. It is evident that abstract q-state $|\Psi\rangle$ does not depend on any operations upon I-frames. All symmetries would enter via the wavefunction $\Psi(\mathbf{x}_1,…,\mathbf{x}_n)$ or functional there off, viz. a density functional. With the number of 3n classic degrees of freedom, semi-classic models may follow; elementary matter sustaining q-states enter as parameters feed in by using the order introduced with configuration space. Quantum degrees of freedom enter via quantum numbers related to self-adjoint operators and their eigen-states: spin operators and 2- and 4-spinors basis functions are well known examples.

**Proper semi-classic models**
In early semi-classic models were coordinates customarily referred to particle positions, e.g. electrons, nuclei,etc. The (real) numbers {**x**} could be loaded with a meaning (that is to be eventually associated to classical particle positions for instance) though that does not affect practical operations (computations). It goes without saying: it is not required by abstract mathematical structures, because no classical physics representation is sought; moreover remember, both domains are non commensurate.

Labels 1-to-n can be traced to particular elementary material elements; the choice allows use of invariant ordered configuration space. For identical elements a second sub-index facilitates handling of permutations symmetries, see POL. [5]

With respect to I-frames one can distinguish internal and external q-states; these latter require another particular laboratory reference frame thereby opening handling for classical physics simulation models as well. [15] Yet gauge symmetries [6] incorporated in the theory can function via q-numbers (irreducible representations). In what follows we focus on quantum numbers for the internal as well as for electromagnetic (EM) model system where a (laboratory) source would either define an origin or signal a direction target for the emission case.

**Double-slit case: how does q-scattering work?**
To help appraisal a notion of q-state, consider a laboratory double-slit experiment such as the one reported by Tonomura [16] but viewed now from frameworks where elementary materiality (electrons) do not play a classic physics role, yet only as executive support of q-states (this means ensuring functionality). The configuration space covers all infinite values related to the presence possibilities though of an executive kind that exclude locations in space as classical elements. *Electron states* are basic elements, supplemented by detection devices. It is an interface where q-events may enter stage.

Tonomura developed technology so that the equivalent of "one-electron-at-a-time" was present in the region supporting a device comparable to a double slit. [16] Actually, the statement must be attuned: *only one executive presence at a* time. This would be enough to sustain a q-state independently of the double-slit material device. A key point: the same q-state necessarily would include response to interaction zones with detection screens; now, the quantum label **k** indicates direction orthogonal to the surface sustaining the double slit that would act as *q-scattering centers*.

q-Scattering is a relevant component for the quantum theoretical analysis; only its executive presence is required to the extent that only one q-event at a time is intended; yet *the q-state covers full configuration space independently from presence of elementary materiality*. It is the interaction between a q-state and a double slit that counts, and the result would prompt all *possibilities* associated, not the sustaining material at a given time and location; these latter terms belong to classical physics.

The scattered q-state by hypothesis would show spherical symmetry, each at one slit's center acting as origin and this because it concerns all scattered q-possibilities *at once*; any other symmetry determined by the q-operator associated to the slit being reflected as well. Note that the meaning with executive presence hint at no object interaction required for the effect to occur: only a scattered q-state of local spherical nature and therefore without further ado *no trajectories involved*. [9,14] This is the nature of quantum effect intended as scattered q-state. A full-blooded mystery in a classical physics world. Yet it is a well-posed quantum case.

In other words, a q-state *interacts* with q-states for the double-slit including an interaction operator technically giving relevant laboratory geometric information. *It is neither a particle nor a path that are the relevant features*. It is q-*interaction* that is to be explored: first, from an abstract standpoint thereby skimming all possibilities; second, semi-classic elements (operators related to laboratory locations) enter to complete the narrative. All these elements make plain an executive presence: production of functionality although not worried by classical particle presence since *QM does not describe the materiality whereabouts*. [9,13,14]

Incidentally, it goes without saying, this view is at variance with standard QM. It is a radically different one.

The scattered q-state presents coherence (by construction) and at a given distance via q-interaction interference effects develops; this scattered q-state grants two q-state components that can be voiced as "propagating" in a mathematic space (both forward and backwards). Again *the whereabouts of the elementary materiality supporting the q-state* are not part of the problem. All possibilities are taken in.

It is at this point where one can weigh up another fundamental difference quantum and classic views have on the issue. All classical language must be suppressed when addressing the quantum domain otherwise inevitably weirdness occurs. [14,21] The quantum formalism should suffice: In a proper semi-classic model it predicts an (possible) interference pattern! And, of course, one cannot understand the quantum with the classic: they are non-commensurate. Such predicament is new since now what is under scrutiny corresponds to *measurement as a physical process*.

In summary: Quantum states are first handled in Hilbert-Fock space, not in real (laboratory) space, thence interaction with a double slit device is given a q-operator form (in the form of a scattering potential); these interaction centers have associated the scattered q-states giving sets of new possibilities embodied in the final scattered q-states: |slit-1> and |slit-2> projected at each slit. [9] These new q-states open over new possibilities. Once you understand that a full set of new possibilities is involved (and not the passage of particles), the interference pattern is a natural mathematic result, in particular, for one elementary sustaining matter devices. Questions such as: which way took the photon or the electron are meaningless; they do not belong to a q-description. This latter question is classical world problem not accountable by quantum schemes. Thus one should not use quantum stories to explain classical situations without further ado, great care must be exercised.

**Photonic framework view**
In abstract space all possibilities should be included, so that the result (in this case) is a q-state signaling a possible interference pattern. This situation (in abstract space) can be set in conformity (linked) to a putative "simultaneous" interaction at double slit device of *one and the same incident q-state*, (thus simultaneity is trivial). This introduces a coherence feature. The experimental setup grants this sameness. Interposing a *detecting* surface beyond the double slit location, and in agreement to (semi-classic) calculations, an *interference pattern* should *emerge* from actual q-events recorded there (viz., remind $(\varepsilon_i-\varepsilon_j)$ |i>⊗<j|q-state> may link to a possible event structure); and be patient, until a sufficient number of q-events are being collected. Actually it does as predicted with q-physical tools. [21] Experimentally the case presents as follows.

Detection of electron-states one-by-one apparently generates random clicks during initial collection of spots (clicks). However, according to present perspective (sustainment), any two q-events will be independently *correlated* by interaction at the double slit via the final quantum state containing *full information* on q-interactions (<j|q-state>-term); the arrival time does not play a role (this one again is a classical physics element). Yet, any sustaining material would be a "carrier" for the *same* q-state though the *prediction* of single spot localization is not legitimate. No quantum scheme predicts localization of a q-event. No independent particles model can do it either; we are thence in presence of q-interactions at the double-slit (or its equivalent) yielding q-states in interference scheme. The abstract interference pattern will be there anyway, first as calculation possibility, and, then at laboratory until enough q-events are recorded an interference *image* pattern would slowly emerge. The classical physics paradigm has been replaced by a quantum one and words such as "impinge" are not recommendable as they convey a wrong message. There is nothing weird here now. Yet what happens in between is not a quantum concern it is one that make sense for the description in classical physics besetments. And this latter is the source of weirdness.

It is a q-event that can engender a "click"; this one would parallel to real quantum energy/ momentum transfer linking the q-state to the one on the functional surface; there are no flying q-events (no trajectories of particles). The sensitive surface records q-events that, initially after collecting a few of them, appear *as if they were random in location*. The event obviously includes the detector screen as well (not emphasized).

As noticed by Tonomura [16], early events do *appear* as being quite randomly distributed; this is the impression at least and it suggests that a too positivistic *interpretation* misses the physics encapsulated in a q-event. Gathering q-events in an separate device and once all of them are simultaneously exposed one would get an *image* that emerges as a result highlighting the underlying q-state one wanted to probe.[9] The three elements in $(\varepsilon_i-\varepsilon_j)$ |i>⊗<j|q-state> are not separable nor distinguishable in the q-event materializing it.

**q-Information flows**
Reading quantum states is one way to ascertain physical and/or physical chemical events. Besides the q-event at sensible surfaces may put in action a detector *qua* quantum device; the q-state now, translated into an entangled state, becomes certainly not equivalent to a classical apparatus. For instance the interference image hides so to speak the underlying supporting materiality of the detection events: namely, information "flow" from the q-state to the detector q-state. This latter element is not representable in a classic sense; one can gist an "opening" state and then a "closing" state. Thus a q-state should also include basis states corresponding to the detector system: this implies a *contextual* situation. The associated "suture" is to be included in the q-interaction. This aspect is elaborated to a certain extent in what follows.

Once adopted for the present framework not a trace of classical physics elements remains as foundations of quantum measure theory.

Information is the "stuff" conveyed by quantum processes. [14,18] This can be read from sets of q-states as illustrated now with an example.

**Opening access to spin triplet states**
This case import to the extent that it shows a quantum physical way to intersystem crossing; in the standard semi-classic approach one employs potential energy curves (profiles). We take advantage now from the experience gained in reading q-states.

A relevant base vector in this case with obvious simplifications reads:

$$(|i=0\rangle \otimes |1_\omega\rangle \ldots |i=0;1_\omega\rangle \otimes |0_\omega\rangle \ldots |i=1\rangle \otimes |0_\omega\rangle \ldots |i=1;0_\omega\rangle \otimes |0_\omega\rangle \ldots |i=2\rangle \otimes |0_\omega\rangle \ldots) \quad (5)$$

Consider a start up q-state to help develop the q-arguments:
$$(C_{i=0 \otimes 1_\omega} \ldots 0_{i=0;1_\omega \otimes 0_\omega} \ldots 0_{1;0_\omega} \ldots 0_{1 \otimes 0_\omega} \ldots 0_{2,\otimes 0_\omega} \ldots) \quad (6)$$
From this q-state only ground closed shell state ($|i=0\rangle$, $S_0$) may act as root state, and the lowest electronic excitation state $S_1$ can directly be open therefrom. Yet the requested target q-state looks like:
$$(0_{i=0 \otimes 1_\omega} \ldots 0_{i=0;1_\omega \otimes 0_\omega} \ldots 0_{1;0_\omega \otimes 0_\omega} \ldots 0_{1 \otimes 0_\omega} \ldots 1_{i=2,\otimes 0_\omega} \ldots 0 \ldots) \quad (7)$$

To move amplitudes in (6) to attain (7) requires a number of q-physical operations. The base state sustaining $1_{i=2,\otimes 0_\omega}$ corresponds to a spin triplet excited electronic state. And let us test the following procedure: From amplitude at excited electronic base state $|i=1\rangle \otimes |0_\omega\rangle$ that originates from an electronic excitation process from $S_0$, Cf.(6) only a spin singlet obtains.

Now a process, similar to optically activated zero-field magnetic resonance phenomena [19,20] may be evoked to progress via coherent photon-matter states. Thus, q-state using the basis (5) a first entanglement brings to:
$$(C_{i=0 \otimes 1_\omega} \ldots C_{i=0;1_\omega \otimes 0_\omega} \ldots 0_{1;0_\omega} \ldots 0_{1 \otimes 0_\omega} \ldots 0_{2,\otimes 0_\omega} \ldots) \quad (6')$$
This case heralds first a photon state *entangled* with ground state basis. Both amplitudes can be different from zero simultaneously; energy conserves if relation $|C_{i=0 \otimes 1_\omega}|^2 + |C_{i=0;1_\omega \otimes 0_\omega}|^2 = 1$ holds. The very initiation process relates then to entanglement involving one-photon state and ground state, (6'). This looks like a "suture" of a free photon base states to the material-sustained bases (see below for details).

Yet the amplitude $0_{2,\otimes 0_\omega}$ (at triplet base state) has so far no information that would conduct the system along the spin triplet base state where the "local" spectrum would concern $\omega'$, a low frequency one. To open it a one photon states $|1_{\omega'}\rangle$ must be drawn first from an external source and interact with the q-state. A base state in a photonic q-state affected by a zero-valued amplitude cannot respond to an external probe. Thus use of the S-T *energy gap injection* as photon states with appropriate resonant frequency that may turn out to be a key to the opening a channel towards setting non-zero value amplitude at the triplet base state. Thence a second resonant photon-state would lead to a $(1+1)\hbar\omega'_{ST}$ (induced) emission cascade resulting in the triplet state activation. Measuring the process from the spin-singlet ground electronic state two units of angular momentum were necessary. The role of two S-T gap energy quanta legitimates conservation of angular momentum.

**Gangways/gangplanks**

Let us now examine with the required formalism to link independent partite states to a coherent state embodying the equivalent of the total elementary sustaining matter. Such sutured q-state might reflect either entrance or outgoing channel from/to a gangway state. More detailed: one starts up from separate matter and radiation sustained system states: |rad-sustained> ⊕ |matter-sustained>. A one- photon radiation base state looks like:

$$|1\text{photon base state}> \to (|0_\omega>\ \ |1_\omega>) \qquad (8)$$

It has q-bit structure. One can see that vacuum information would always be present. Now consider independent partite states [21]

$$(0_{|0_\omega>}\ 1_{|1_\omega>}) \oplus (1_{i=0}\ldots 0_{i=1}\ldots 0_{i=2}\ldots). \qquad (9)$$

In this example the matter-sustained q-basis has not yet registered information on the photon q-field as it were. This information cannot be conveyed by ⊕ operation. It is a direct product form (⊗) that stands for such interaction information. To elicit it let start with direct product opening for interaction:

$$(0_{|0_\omega>}\ 1_{|1_\omega>}) \otimes (1_{i=0}\ldots 0_{i=1}\ldots 0_{i=2}\ldots). \qquad (10)$$

On the trail of interaction the momentum being traded between the partites with the following caveat: the q-state sustained by elementary materiality would now include the incoming information (a way to acknowledge interaction) and the "suture" state:

$$(0_{|0_\omega>}\ 0_{|1_\omega>}) \otimes (0_{i=0\otimes 1_\omega}\ldots 0_{i=0;1_\omega\otimes 0_\omega}\ldots 1_{1;0_\omega}\ldots 0_{1\otimes 0_\omega}\ldots 0_{2\otimes 0_\omega}\ldots 0\ldots). \qquad (11)$$

It is the q-state signaling a vacuum $(0_{|0_\omega>}\ 1_{|1_\omega>}) \to (0_{|0_\omega>}\ 0_{|1_\omega>})$ information transfer; this q-state defines a "gangplank".

The vacuum state information enters as labels re-normalizing the base set. And it would remain as labels in the elementary materiality q-state. The photon and matter sustained basis at information level appear now entangled. The latter gangplank state is normally over-looked. Yet it is essential to invert the process in the direction of photon emission.

The eqs. (10), (11) can be seen as forming a gangway to traffic photon states with a matter-sustained q-field.

Second, examine "propagation" from state (6) over base states made accessible by a *second* one-photon state interaction (the corresponding gangway being implicit). Focus attention on fluorescent-like states: from (6) as pivot, to get at a triplet state activation one may follow a course that signals laboratory actions: 1) Information-injection of $S_1$-$T_1$ *energy gap* via a photon state say ($|0_{\omega'S1-T1}>$   $|1_{\omega'S1-T1}>$) and q-state ($0_{|0_{\omega'S1-T1}>}$  $1_{|1_{\omega'S1-T1}>}$); and 2) label entanglement suggested by state (7) below identified by amplitude with appropriate notation $C_{1*;0_\omega \otimes 1_\omega ST}$:

$$(0_{i=0 \otimes 1_\omega} \ldots 0_{i=0;1_\omega \otimes 0_\omega \otimes 1_\omega 'ST} \ldots C_{1;0_\omega \otimes 0_\omega 'ST} \ldots C_{1*;0_\omega \otimes 1_\omega 'ST} \ldots 0_{2, \otimes 0_\omega \otimes 0_\omega 'ST} \ldots) \quad (12)$$

Note two features: 1) the gangplank ($0_{|0_{\omega'S1-T1}>}$  $0_{|1_{\omega'S1-T1}>}$) vacuum q-state is implied for the time being to alleviate notations; 2) the addition of an excited state base state $|1*;0_\omega \otimes 1_\omega ST> \otimes |0_\omega>$. This later by construction has two pieces of information i.e. energy level above the first excited state ($S_1$) and the energy level when measured from ground state ($T_0$) displays the equivalent of $(1+1)\hbar\omega'_{ST}$ so that the triplet base state label now gains necessary information along the new channel named as ($S_1$-$T_1$). And energy conservation demands inclusion of a second external photon state at frequency $\omega_{ST}$; this one would "harvest" an assisted consecutive two photon induced emission that would result in the triplet state opening via a non-zero value for the matching amplitude:

$$(0_{i=0 \otimes 1_\omega} \ldots 0_{i=0;1_\omega \otimes 0_\omega \otimes 1_\omega ST} \ldots 0_{1;0_\omega \otimes 0_\omega ST} \ldots 0_{i=1*;0_\omega \otimes 1_\omega ST} \ldots 1_{i=2, \otimes 0_\omega \otimes 0_\omega ST} \ldots 0 \ldots) \quad (12')$$

Observe that a direct one-photon activation from singlet ground state $S_0$ to the triplet $T_1$ is not allowed due to AM-conservation rules *unless* the subsidiary photon state *also displays by special preparation orbital momentum space OAM* [23] but, at this point, this case is no further considered.

Once the activation channel displays non-zero amplitude at state (12') only the first excited state $S_1$ could have act as root state for the *supplementary* electronic excitation events. And a *non-rotating* wave model supplies energy at $2\hbar\omega'_{ST}$ that also *add label information* at position $0_{2,T \otimes 0_{\omega'}}$; now the base state can play the role of a dressed vacuum, i.e.: $0_{i=2,T \otimes 0_\omega \otimes 0_{\omega' ST}}$. Standard rotating frame models are not useful in this context simply because they have a too classical (electrodynamics) sight.

Thus, shine a *second* ST-gap photon state $|1_{\omega ST}>$ at (11) to prompt for the cascade noted above; this would end up with non-zero amplitude at $C_{|i=2,S=1> \otimes |0_\omega> \otimes |0_{\omega ST}>}$, these states remain implicit in the notation of (12') now update into the more explicit sought result:

$$(\ldots 0_{i \otimes 1_\omega} \ldots 0_{i;1_\omega \otimes 0_\omega} \ldots 0_{i+1;0_\omega \otimes 0_\omega 'ST} \ldots 1_{i+2,T \otimes 0_\omega 0_\omega 'ST} \ldots) \quad (12')$$

The information brought up by injection and registered by the second ST-photon gap state would add amplitude shifting from $0_{i+2,T \otimes 0_\omega 1_\omega ST}$ into $1_{i+2,T \otimes 0_\omega 0_\omega 'ST}$, as shown above the operation shares a flavor of information supplement via labels (prompted by the physical injection). One is far from the spin-orbit levels crossing view.

In other words, chemical and photo-physical processes starting at a triplet as root state can now begin. The ST-gap two photons pay for two units of angular momentum required by the "transfer" from the spin-singlet electronic excited state. The space part corresponds to changing from L=0 to L=1. [23] The supporting elementary materiality remains unchanged in numbers, only the q-state changes.

The elementary event that may lead to a transfer of one energy quantum could have taken (10) as portal state. *Mutatis mutandis* this type of state may also act as possible portal for photon emission that includes the proper gangway q-state. Thus, *information would circulate expressed via q-state changes in amplitude*. And this is a main result that closes the elementary examples chosen to illustrate the photonic q-model.

**Decoding QM from Photonic Clues**
The possibility for a re-foundation of quantum mechanics (QM) is now open provided quantum states (q-states) do not represent objects (entities). This is the necessary condition. To make a long story short take the complete set of proper base states $\{|i\rangle\}$ as being sustained by the elementary materiality that is the case. The amplitudes can open/close access to root states. In short the amplitudes control access to possible q-interactions.

By "ignoring" the q-photon fields, intersystem entanglements are left outside. Clearly standard QM alone cannot be a complete framework. There is no physical way for information transfers. Wigner's friends lead to nonsense.

The standard formulation of QM supplemented with a non-representational mode would be an appropriate semi-classic framework with the following caveats:

Spaces paved by possibilities displace those grounded in probability. For the case where only one elementary materiality at a time acts via q-scattering two elements are always determinant: 1) trajectories are never a way to describe the system, it is the q-state that matters; 2) q-scattering implies all possibilities opened and determined by the nature of the scattering center expressed as scattered q-states. In few words, there is no way to set up the situation with classical physics tenets even less to understand in classical physics words any quantum physical effect. Presentations of semi-classic models may use intermediate devices and language adapted to this situation.

The resulting abstract form once linked to chemical semi-classic models leads to a suitable quantum Mathematical Chemistry. In the abstract scheme there are no material objects, only abstract quantum states and self-adjoint operators: basically a Hilbert space formulation. One great virtue of the standard Hilbert space formalism of QM is that its outcomes concern variability meaning with this that when repeating twice the same experiments, it will generally produce a different outcome among all possibilities available. Tonomura's experiment discussed above showed this behavior type. But then, the effect of the q-state at laboratory space might show up as an image after enough registered q-events are collected. The possibility space is in principle complete yet the one that is related to the matter-sustained q-states only is not necessarily complete.

Projecting from abstract to laboratory space requires I-frames and concomitant introduction of configuration space coordinates. These coordinates prompt inclusion of rigged Hilbert spaces to help decode abstract vectors e.g. wave functions. These coordinates can also be mutated into labels and thereafter a metric form introduced. Dynamical variables are turned into Hermitian operators in Hilbert space as usual. Linkages between abstract and laboratory space are made more explicit.

Naturally, scattering and interference phenomena also appear in classical settings, however, *discovery of interference in a quantum setting does not logically imply a wave property of the carrier in a classical sense.* Such conclusion would not be logically granted. And this paper suggests that it is quantum physically a non-sense. Quantum physics is without objects only q-states are in. [9,14,18,25]

Quantum mechanics then with the assistance of possibility spaces concept become now a differently grounded scheme not directly attached to positivistic tenets. This is a significant by-product of the photonic approach.

**Discussion**
This paper gives preeminence to abstract q-states, however, using the concept of configuration space (I-frames), geometry factors would enter both via measure elements (integrals) and self-adjoint operators and then, introduction of invariance groups leads to particular quantum number families. In this manner via semi-classic models one approaches a blend of Mathematical Chemistry and Physics with an important caveat.[25,26] The elementary materiality plays a role of executive (functional) presence instead of presence as objects. The shift in ontology grounds is decisive to eliminate old disputes between idealists and realists that rages even today. A sort of reality without realism takes over the grounds where mathematical formulations would evolve. Registered spectral changes will uncover matter-sustained processes.

If we call a register of a q-event an outcome of a measurement it is clear that it is determined by physical interaction between measured and measuring elements (laboratory space) yet not by an "observer". Therefore it would never be determined by internal variables of the q-event only.

The grounds on which the photonic view differ significantly from the standard ones can be gauged with the help of a thorough knowledge of the standard framework; to cope with this the reader may consult a paper by Plotnitsky [27] where one finds a comprehensive description of foundations of quantum physics in the Copenhagen spirit. The differences noted by us rest on a new floor. Some differences are summarized: 1) It is a non-representational framework; 2) The key concept is given by Planck discovery that energy and momenta *exchanges* in finite amounts; 3) Energy levels are a resource not a property of a quantum system, so the elementary constituents do not necessarily occupy such levels; 4) the numerical values obtained by bringing the Hamiltonian to diagonal form do not represent such energy levels, only the difference between two may signal a finite amount of energy; 5) scattering imparted by a particular interaction operator will always produce a swarm of possibilities, even if at laboratory one may prepare supportive elementary material one-by-one, the handling in theory would always produce a full fledged scattered q-state; 6) The q-state does not represent the whereabouts of the elementary materiality, this latter *sustains* (supports) the physical q-state in a way not describable in classical world terms; 7) use of Schrödinger Hamiltonian permits obtaining information on eigenvalues and introducing asymptotic boundary conditions sets of eigen-functions obtain; 8) use of Dirac operator (4x4) matrix produce information on eigenvalues and spinor eigen functions under particular boundary conditions, connections to Schrödinger non-relativistic equation are found in Feynman´s book [28]; 9) inclusion of quantized fields linked to electromagnetic energy completes the presentation of quantum dynamical systems including interactions; 10) Matter and electromagnetic systems relate in the form of entangled non-separable systems; 11) Intersystem crossing is given a physical q-description; 12) Classic epistemology is not relevant to understanding quantum physics, this latter should generate a sort of q-epistemology focused on a quantum world.

The above are some relevant issues concerning quantum systems related to charged and finite non-zero rest mass. Concepts of classical particles and waves have to be replaced and taken away from our thinking thereby letting take the place to renewed forms relying on mathematical frameworks. [28,29] Conditions to link low energy region to high and to very high energy ones are met by this non-representational mode. Thus including non-commutative geometry and algebraic theory of spinors with Clifford Algebra [24,28,29] important examples can be worked out that introduce physics at energy regimes going well beyond present low-energy phenomena. [24,30]

In closing the paper it is quite instructive to give a somewhat long citation from the preface to the first edition of "The Principles of Quantum Mechanics" by Dirac: [31] *The classical tradition has been to consider the world to be an association of observable objects (particles, fluids, fields, etc.) moving about according to definite laws of force, so that one could form a mental picture in space and time of the whole scheme…Her* (Nature) *fundamental laws do not govern the world as it appears in our mental picture in any direct way,…* [31]

Yet this fundamental view advocated by Dirac was forgotten. Our paper is a self-effacing contribution to formulate a new ground for Quantum Mechanics that incorporate in a way those ideas although going beyond the representational mode.

Both measured and measuring devices comprise quantum elements that might prompt for q-events. The quantum system can be subjected to q-interactions that do not imply detectable energy swap also; this could be the case (at least in part) with virtual double-slit device or similar as analyzed here whenever the interaction does not produce transfers of energy quanta measure by final reckoning of amplitudes concerning starting and end point.

If one takes "reality" to be that one of "clicks" appearing in a particular experiment then, according to the present approach, they hide more than they uncover. In particular, correlations induced by the wavefunction that the collected information would put in evidence once the associated image becomes apparent; but, as possibilities, they are always there enabling the characteristic quantum behavior. So, all individual q-events (to recover our formulations) are not innately random yet locations separately taken are unpredictable in terms of laboratory space and time coordinates.

An interesting corollary follows: There is no need for an Earth based laboratory to produce chemical changes in interstellar space chemistry.

**Acknowledgement**

The author is indebted to Prof. E. Ludeña for cogent and enlightening discussions.